\documentclass[useAMS,usenatbib]{mn2e}
\usepackage{rotating}
\usepackage{journals}
\usepackage{graphicx, subfigure}
\usepackage{xspace}
\usepackage{amssymb}

\def\farcs{\hbox{$.\!\!^{\prime\prime}$}}
\def\degr{\hbox{$^\circ$}}
\def\arcmin{\hbox{$^\prime$}}
\def\arcsec{\hbox{$^{\prime\prime}$}\xspace}

\def\kms{$\rmn{km\,s^{-1}}$}

\def\ang{$\rmn{\AA}$\xspace}

\def\chan{{\it Chandra}\xspace}

\def\src{CX93\xspace}
\def\tmp{CXOGBS\,J174444.7$-$260330\xspace}

\newcommand\msun {M$_{\odot}$\xspace}
\newcommand\rsun {R$_{\odot}$\xspace}

\newcommand{\ergsec}{$\rmn{erg\,s^{-1}}$}

\LARGE \normalsize \title[The mass of the compact object in \src] {\tmp: a new long orbital period cataclysmic variable in a low state\thanks{data from ESO programs 085.D-0441(A) and 087.D-0596(D)}}

\author[Ratti et al.]  {E.M.~Ratti$^{1}$\thanks{email : e.m.ratti@sron.nl}, T.F.J.~van~Grunsven$^{1}$, P.G.~Jonker$^{1,2,3}$, C.T.~Britt$^{4}$, R.I.~Hynes$^{4}$, \newauthor D.~Steeghs$^{5,2}$,    S.~Greiss$^{5}$, M.A.P.~Torres$^{1,2}$, T.J.~Maccarone$^6$, P.J. Groot$^{3}$, C. Knigge$^{6}$, \newauthor
 V. A. Villar$^{7}$, A. C. Collazzi$^{4}$, V. J. Mikles$^{4}$, L. Gossen$^{8,4}$ \\
$^1$SRON, Netherlands Institute for Space Research, Sorbonnelaan 2, 3584~CA, Utrecht, The Netherlands\\ 
$^2$Harvard--Smithsonian Center for Astrophysics, 60 Garden Street, Cambridge, MA~02138, U.S.A.\\
$^3$Department of Astrophysics/IMAPP, Radboud University Nijmegen, P.O. Box 9010, 6500 GL Nijmegen, the Netherlands. \\
$^4$Department of Physics and Astronomy, Louisiana State University, Baton Rouge, Louisiana 70803, U.S.A. \\
$^5$Department of Physics, University of Warwick, Coventry CV4 7AL \\
$^6$School of Physics and Astronomy, University of Southampton SO17 1BJ, UK \\
$^7$Massachusetts Institute of Technology, Department of Physics, 77 Massachusetts Avenue Cambridge, MA 02139, USA \\
$^8$Visiting astronomer, Cerro Tololo Inter-American Observatory, National Optical Astronomy Observatory, which are operated by the Association of Universities for Research in Astronomy, under contract with the National Science Foundation. \\
}

\begin{document}

\maketitle

\begin{abstract}  

We present phase-resolved spectroscopy and photometry of a source discovered with the \chan Galactic Bulge Survey (GBS),  \tmp (aka \src and CX153 in the previously published GBS list). We find two possible values for the orbital period $P$, differing from each other by $\sim$13 seconds. The most likely solution is $P=$\,5.69014(6)\,hours. The optical lightcurves show ellipsoidal modulations, whose modeling provides an inclination of  32$\pm1$\degr\,for the most likely $P$.  The spectra are dominated by a K5\,{\sc V}companion star (the disc veiling is $\lesssim$5\%). Broad and structured emission from the Balmer lines is also detected, as well as fainter emission from He{\sc I}. From the absorption lines we measure K$_2=$117$\pm$8\,$\rmn{km/s}$ and $v\sin i=69\pm7$\,$\rmn{km/s}$.  By solving the system mass function we find M$_1=$0.8$\pm0.2$\,\msun for the favored $P$ and $i$, consistent with a white dwarf accretor, and M$_2=$0.6$\pm$0.2\,\msun. We estimate a distance in the range $400-700$\,$\rmn{pc}$. Although in a low accretion state, both spectroscopy and photometry provide evidence of variability on a timescale of months or faster. 
Besides finding a new, long orbital period cataclysmic variable in a low accretion state,  this work shows that the design of the GBS works efficiently to find accreting X-ray binaries in quiescence, highlighting that the spectra of CVs in a low-accretion state can at times appear suggestive of a quiescent neutron star or a black hole system. 

\end{abstract}


\section{introduction}

We present optical follow-up of the X--ray source \tmp, discovered in the Galactic Bulge survey (GBS) of \citet{Jon11}. For the ease of readability, we will refer to the source with its label in the GBS source list, \src. One of the main goals of the GBS is to detect quiescent X-ray binaries (XRBs) - namely binary systems where a white dwarf (WD), a neutron star (NS) or a black hole (BH) is accreting matter from a companion star - that are suitable for dynamical measurements of the mass of the compact object. Those mass measurements can be done through phase-resolved optical spectroscopy of the donor star, provided that good constraints can be put on the binary inclination. If the system inclination $i$ is known, the mass of the accreting compact object can be obtained by solving the system mass function (see, for instance,  \citealt{Cha06}). This requires measuring the orbital period $P$, the semi-amplitude of the radial velocity curve of the companion star K$_2$ and the ratio between the mass of the donor and that of the accretor,  $q\equiv M_2/M_1$. In Roche lobe filling XRBs, with tidally locked companion stars,  $q$ is a function of K$_2$ and of the  projected rotational velocity of the secondary star $v\sin i$ \citep{Wad88}, which is measured from the broadening of the stellar absorption lines.
\newline
Prime targets for this study are eclipsing systems, where the inclination can be derived from the eclipse duration based on geometrical arguments only. Among non-eclipsing systems, the inclination can be estimated by modeling the ellipsoidal variation in the optical lightcurve, caused by the distortion of the companion star shape associated with Roche-lobe overflow (care must be taken if other continuum sources contribute to the lightcurves). 

\begin{figure}
\includegraphics[width=8.0cm, angle=0]{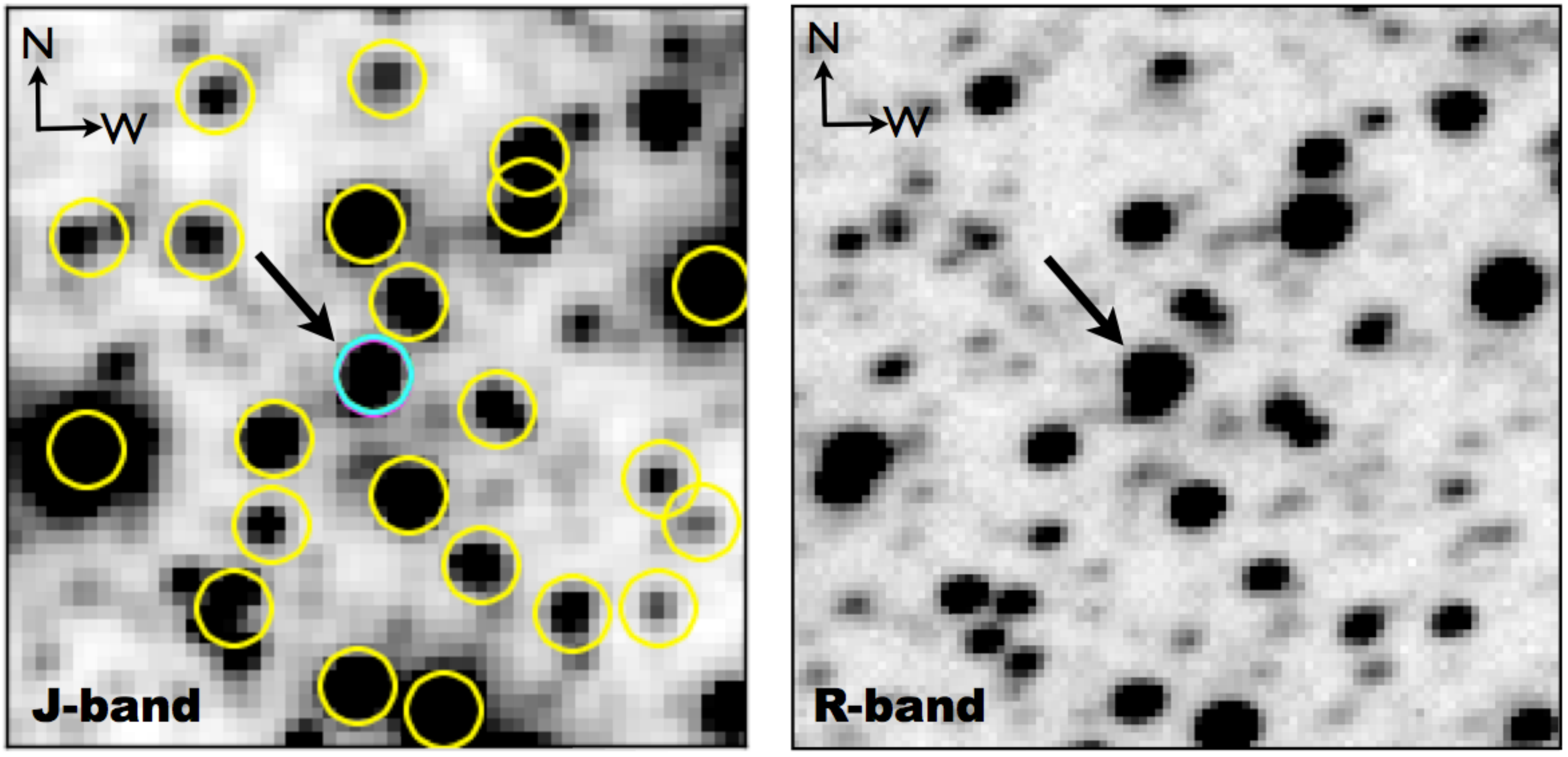} \\
\caption{\label{fig:finders} 20{\hbox{$^{\prime\prime}$}}$\times$20\arcsec finding charts showing the near infrared (J-band) and the optical (R-band) counterpart to \src, indicated by the black arrows. The left panel is from the variable sources in the Via Lactea (VVV) survey, where the VVV source detections are indicated by circles. Two overlapping circles at the counterpart position correspond to the \chan position and to that of the NIR source (R.A. =$17^{\rmn{h}}$\,$44^{\rmn{m}}$\,$44\fs790$, Dec. = $-26\degr$\,$03\arcmin$\,$30\farcs08$, with an accuracy better than 0.1\arcsec). The right panel is an image from the VIsible Multi-Object Spectrograph (VIMOS). }
\end{figure}

Because \tmp was initially not identified as a duplicate,  it has two labels in the GBS source list,  CX93 and CX153 (the offset between the two detections is 3\arcsec). The best X--ray source position, from observation \src, is R.A. =$17^{\rmn{h}}$\,$44^{\rmn{m}}$\,$44\fs791(9)$, Dec. = $-26\degr$\,$03\arcmin$\,$30\farcs3(1)$ (the 1$\sigma$ uncertainties, indicated in between brackets, do not include the 90\% confidence 0\farcs6  \chan boresight error). The source was identified as a quiescent X-ray binary candidate based on preliminary low-resolution spectra of the optical counterpart (Figure \ref{fig:finders}). A near infrared (NIR) counterpart to \src was found in the data from  the VISTA (Visible and Infrared Survey Telescope for Astronomy) variable sources in the Via Lactea (VVV) survey (Minniti et al. 2009, Catelan et al. 2011, Saito et al 2011), with magnitude 14.81$\pm$0.02 in the J-band, 14.01$\pm$0.02 in the H-band and 13.76$\pm$0.03 in the K-band. 
Here we present phase-resolved photometry and optical spectroscopy of the target performed with several instruments. We detected signatures from the mass donor star, which allowed us to measure the orbital ephemeris and constrain the mass of the accreting compact object. 

\section{Observations and data reduction}
\label{sec:data}

\begin{table*}
\addtolength{\tabcolsep}{-3pt}
\caption{ Journal of the observations. The resolution of the spectra is measured from the width of the night sky lines.}
\label{tab:spec}
\begin{center}
\begin{tabular}{llllllllllll}
\hline
Instrument & Date & \# & Exposure   & Grism/Filter & Seeing & Slit  width & Binning & Sp. range & Resolution & Dispersion \\
                    & (UT) &      &    (s)             &                      & (arcsec)& (arcsec)   &                & (\ang)        & (\ang)          & (\ang/px)  \\
\hline
& & & & & Spectroscopy & & & & & \\
\hline
VIMOS & 2011\,Apr.\,29       & 2     & 875            & MR-2.2           & 0.6         &1.0 & 1x1  & 4879-10018 & 10  &  2.50 \\
IMACS & 2011\,Jun.\,23-26 & 4     & 1200-600 & Gri-300-17.5 & 0.8-1.2  & 1.0& 1x1  & 4365-6654    & 4.5 & 1.28 \\
FORS   & 2011\,Jul.\,4          & 4      & 900           & 1200R+93     &  0.9-1.4 & 1.0 & 2x2 & 5957-7280    & 2    &  0.38 \\
MagE   & 2011\,Aug.\,6        & 1      & 900           & Echelle           &  1.2        & 1.0 & 1x1 & 3100-11200  & 1.6 & 0.47 \\
\hline
& & & & & Photometry & & & & & \\
\hline
Mosaic & 2010\,Jul.\,12-18  & 33   &120  & Sloan r$^\prime$ &1.3 & -- & -- & -- & -- & -- \\
Swope & 2011\,Jun.\,27-28 & 360 & 60  & Gunn r & 1.5 & -- & -- & -- & -- & -- \\
\end{tabular}		      
\end{center}
\end{table*}

\subsection{Phase-resolved imaging}
\label{sec:ima}
\src was observed in 2010 Jul. using the Mosaic-2 instrument on the Cerro Tololo Interamerican Observatory 4\,m Blanco telescope (Table \ref{tab:spec}).  
Thirty three 120s-long exposures were collected through the Sloan $r^\prime$ filter. Initial data processing was done by the National Optical Astronomy Observatory (NOAO) Mosaic pipeline \citep{Shaw:2009a}. The lightcurves were extracted using the {\sc ISIS} image subtraction code \citep{Alard:1998a,Alard:1999a}, which provided clean subtracted images.  A variable optical counterpart coincident with the X-ray position of \src was clearly identified, despite the fact the variability is only at the level of $\sim$5\%. We independently analyzed the images using DAOPhot II \citep{Stetson:1987a} and obtained comparable lightcurves, although of somewhat poorer quality than those obtained with {\sc ISIS}.  Since {\sc ISIS} only yields differential count rates, we used DAOPhot II on the best reference image to measure the baseline (non-variable) count rate.  This together with the differential count rates from {\sc ISIS} yielded total count rates as a function of time.  We finally converted these to approximate magnitudes using conversion factors supplied by the NOAO pipeline.  These are deduced by comparison to USNO B1.0 stars, and have an estimated uncertainty of at least 0.5\,magnitudes. The magnitude of \src is 17$\pm$0.5 magnitudes in the r$^\prime$-band.
\newline
\src was also observed using the Direct CCD Camera on the Henrietta Swope Telescope at Las Campanas Observatory on the night of 2011 Jun. 27/28.  A sequence of 60\,sec exposures was obtained through a Gunn $r$ filter for about six hours.  Bias correction and flat fielding was performed through standard CCD data processing in {\sc iraf}\footnote{IRAF is distributed by the National Optical Astronomy Observatory, which is operated by the Association of Universities for Research in Astronomy (AURA) under cooperative agreement with the National Science Foundation.}. {\sc ISIS} was used for image subtraction photometry as for the Mosaic-2 data.

\begin{figure}
\begin{tabular}{l}
\includegraphics[width=8.0cm, angle=0]{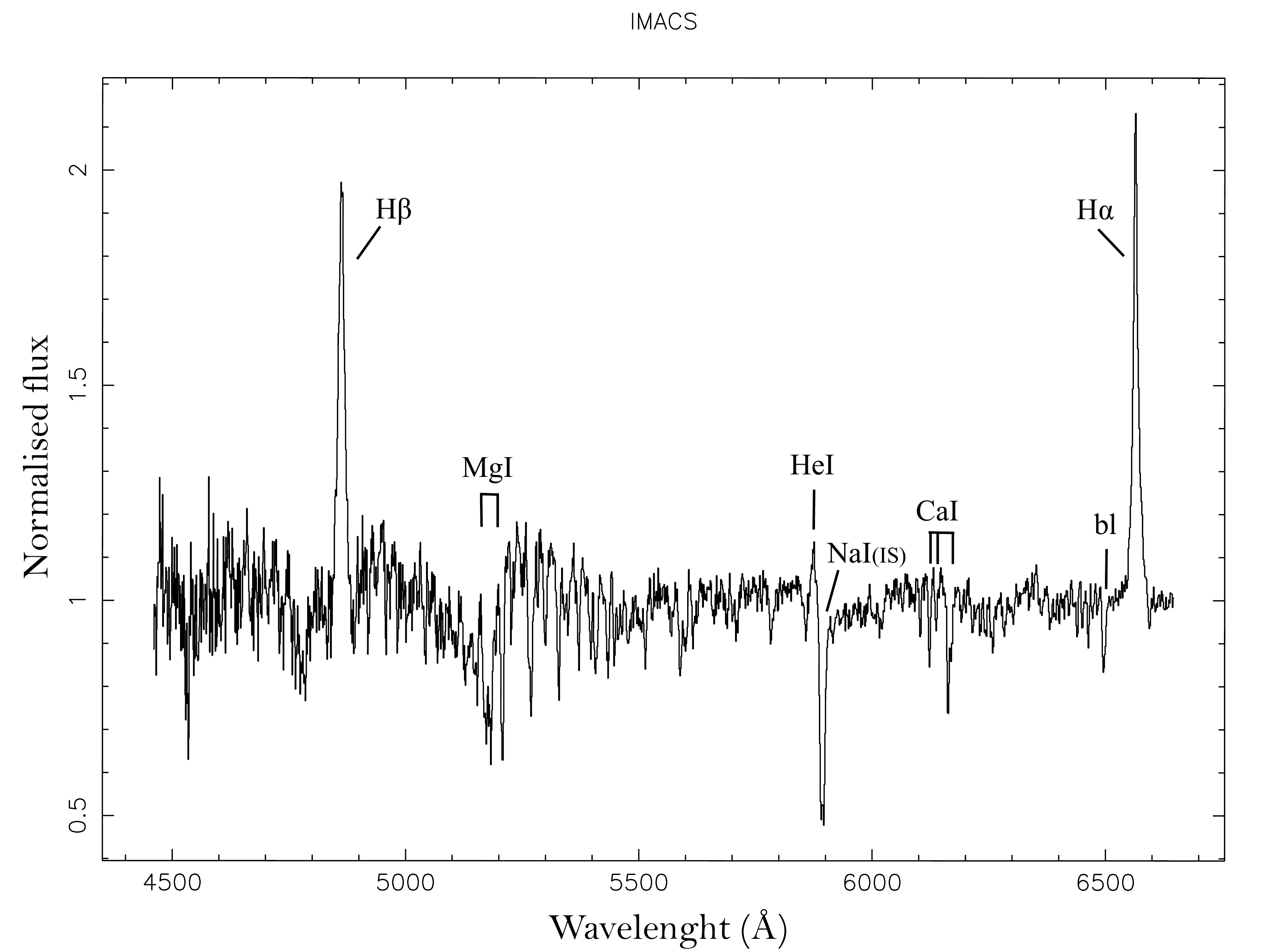} \\
\includegraphics[width=8.0cm, angle=0]{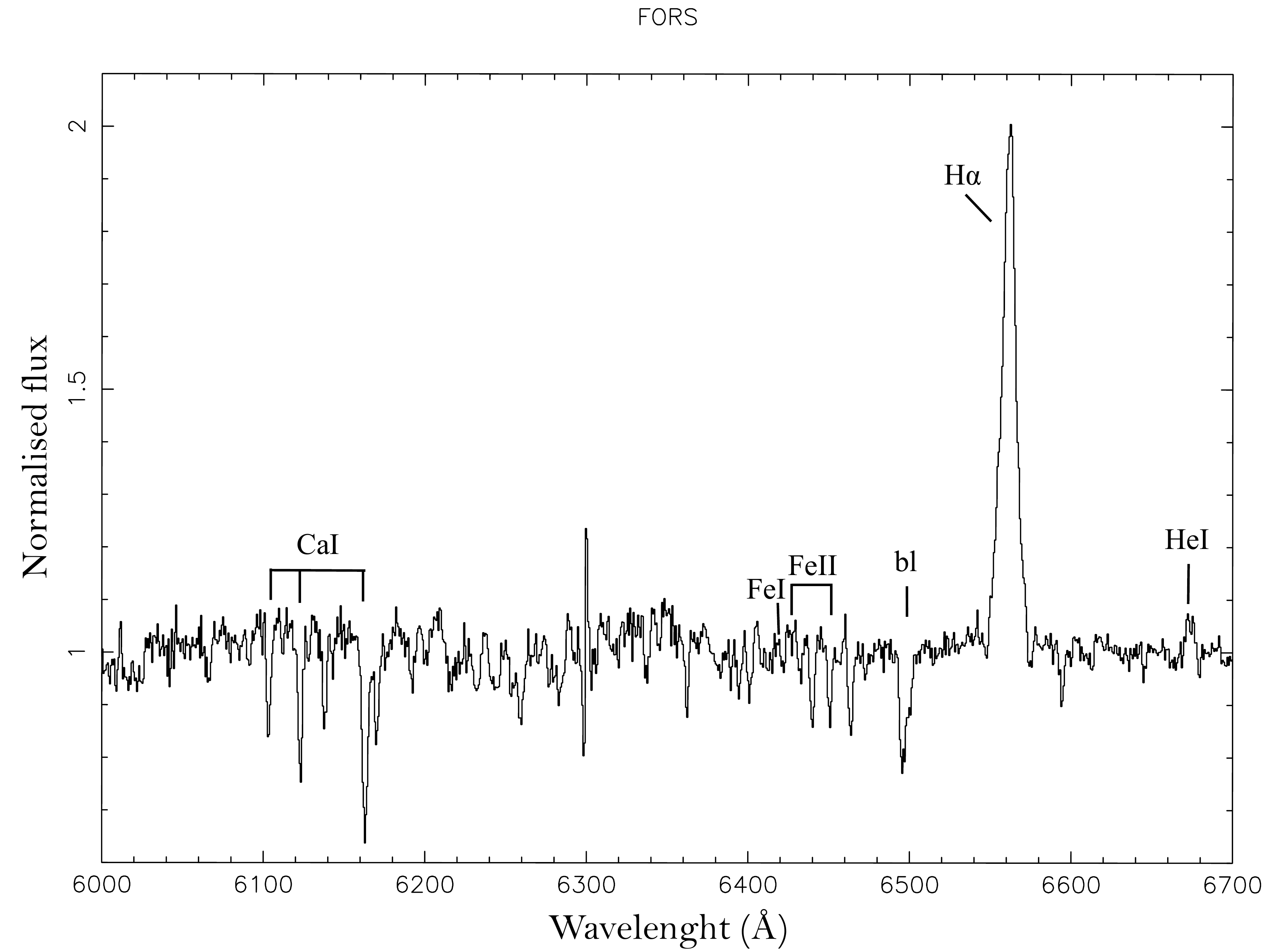} \\
\includegraphics[width=8.0cm, angle=0]{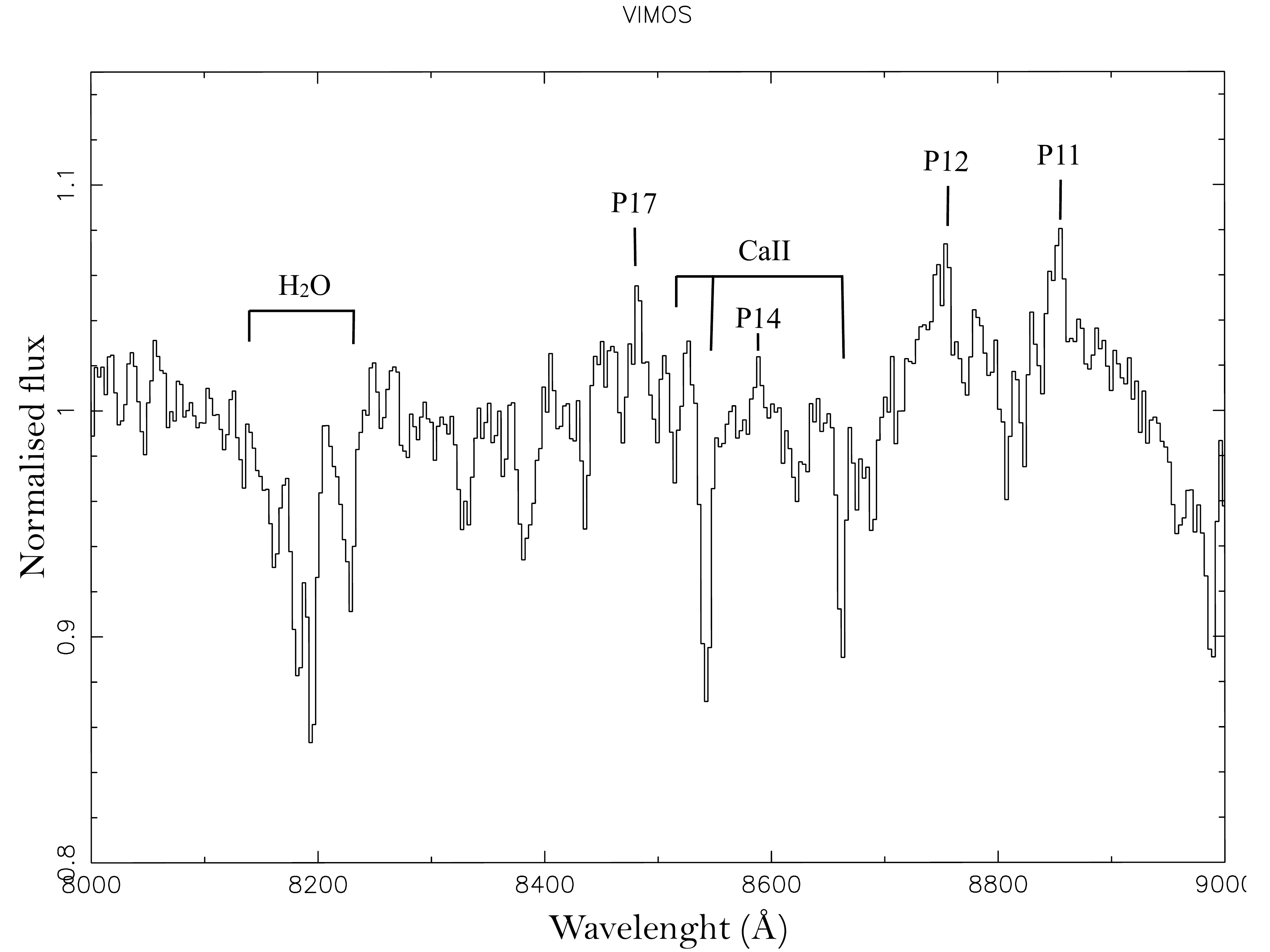} \\
\end{tabular}
\caption{\label{fig:spectra} Main spectral features from \src in the IMACS (top panel), FORS (middle panel) and VIMOS (bottom panel) Doppler-corrected average spectra. }
\end{figure}

\subsection{Optical spectroscopy}
We combined spectroscopic observations from a number of telescopes: the VIsible Multi-Object Spectrograph (VIMOS) and the Focal Reducer and Spectrograph (FORS) at the Very Large Telescope, the Inamori Magellan Areal Camera and Spectrograph (IMACS) at the Magellan telescope, the ESO Faint Object Spectrograph and Camera (EFOSC) at the New Technology Telescope (NTT) and the Magellan Echelle (MagE) spectrograph, also at the Magellan telescope. A list of the observations and the instrument settings is given in Table \ref{tab:spec}. The NTT/EFOSC data are not included in the table and in the rest of this work because of the poor quality of the data due to weather conditions. Template spectra of the stars HD163197 (spectral type K4{\sc IV}) and HD130992 (K3.5{\sc V}) were also observed with FORS and MagE, respectively. 
The images were corrected for bias,  flat-fielded and extracted using the {\sc Figaro} package within the {\sc Starlink} software suite and  the packages {\sc Pamela} and {\sc Molly}  developed by T. Marsh.  We used sky flats for the flat-fielding and we subtracted the sky background by fitting clean sky regions along the slit with a second order polynomial. The spectra were optimally extracted  following the algorithm of \citet{Hor86} implemented in  {\sc Pamela} and wavelength-calibrated in {\sc Molly} with a final accuracy of 0.02 \ang for FORS (using arc exposures  taken during daytime), 0.04 \ang for IMACS, 0.09 \ang for VIMOS and $\sim$0.04 \ang for MagE. In each spectrum, the wavelength calibration was checked and corrected for shifts with respect to the position of the night-sky O{\sc I} lines at 5577.338 \ang and/or at 6300.304 \ang \citep{Ost96}. Each spectrum has been normalised dividing by a spline fit to the continuum, with a maximum order of 10. The full MagE spectrum was extracted, but only order 9 (5630-6673 \ang) and 10 (6265-7415 \ang) were used for our dynamical study of \src, as they cover the spectral region of the H$\alpha$ and Ca\,{\sc I}, which is rich in absorption features for K type stars and where the signal-to-noise ratio in the spectrum is highest. The MagE spectra show a complex continuum, mainly
caused by instrumental response variations across the relevant orders. In order to normalise the spectra,  we clipped two $\sim$300 \ang-long pieces from the order 9 spectrum (5920-6240 \ang and 6310-6630 \ang), and one from the order 10 (6300-6750 \ang). The regions were selected to keep as many absorption lines as possible in one spectrum. By selecting short pieces from the spectrum we could achieve a good fit to the continuum on each piece with a relatively low-order spline (order 10), avoiding as much as possible the risk of altering the equivalent width of the lines that occurs when using high order splines. The overlap between the orders 9 and 10 allows for a double-check of the results. 

\begin{figure}
\begin{tabular}{l}
\includegraphics[angle=0,width=0.45\textwidth]{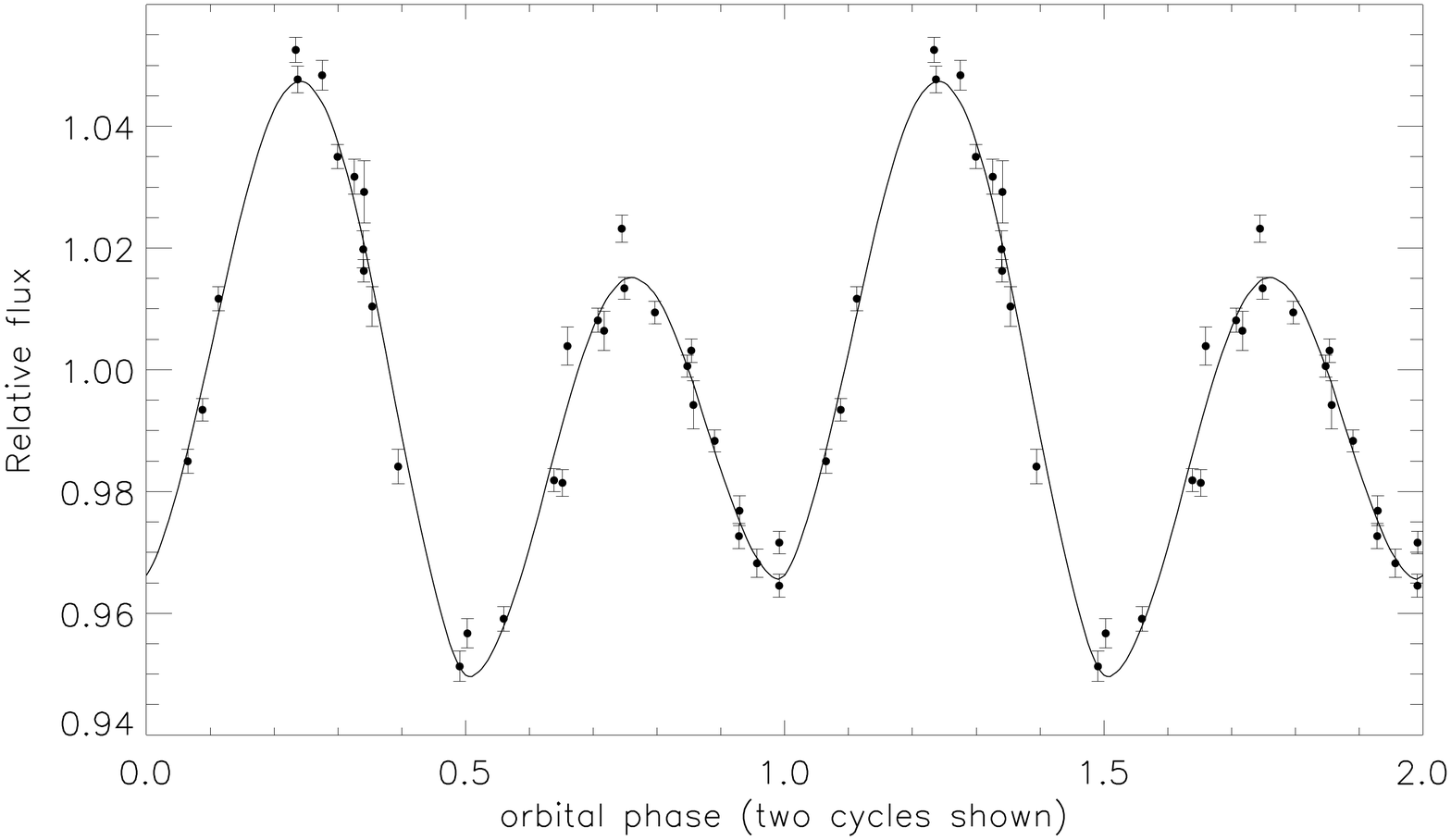} \\
\includegraphics[angle=90,width=0.45\textwidth]{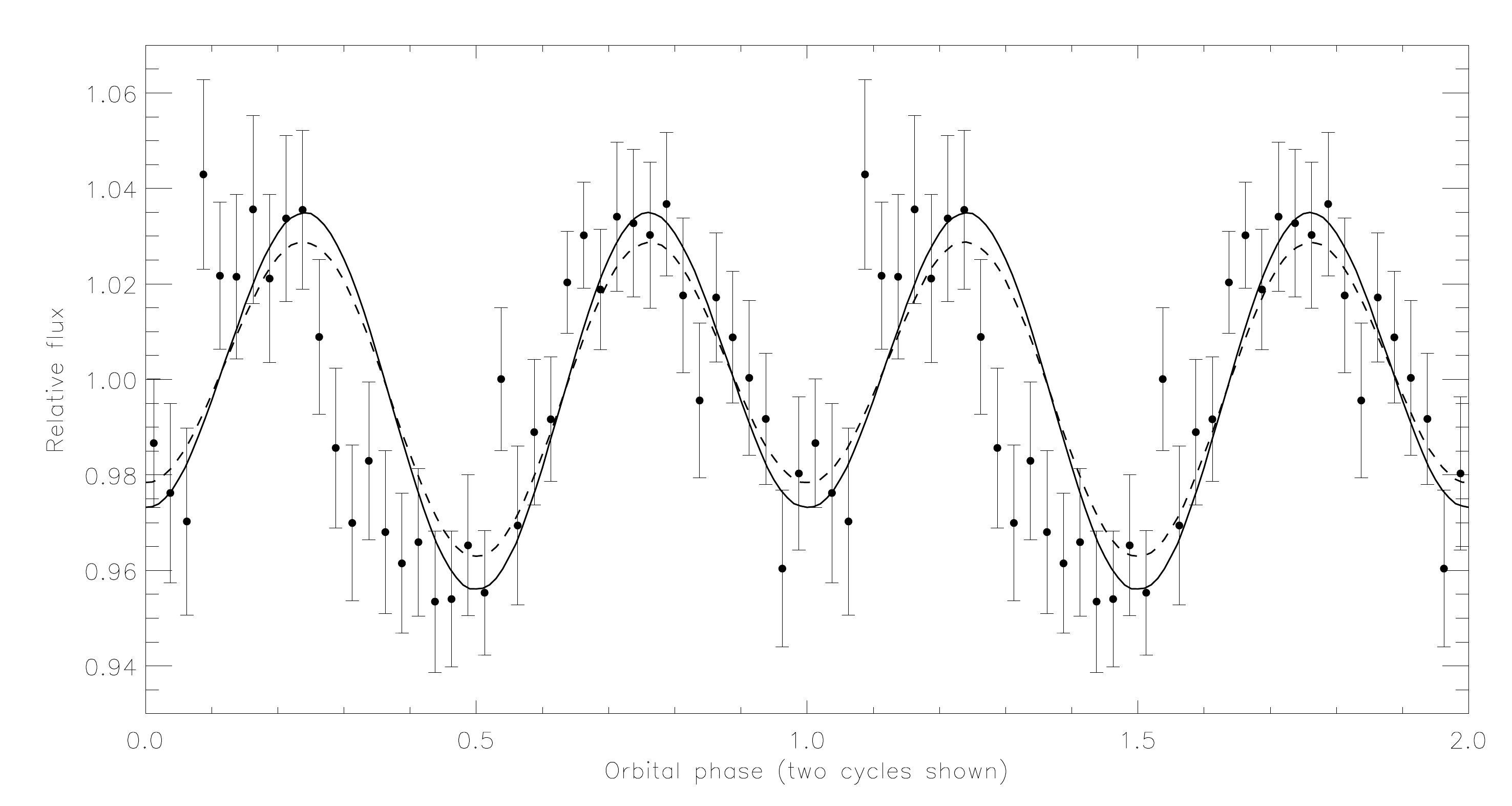} \\
\end{tabular}
\caption{ \label{fig:licus} Top panel: Mosaic-2 lightcurve. The solid line represents the best fit model assuming ellipsoidal variations plus a single bright spot. Bottom panel: LCO lightcurve, fitted with a pure ellipsoidal modulation model for an inclination of  $i=32.3$\degr (dashed line) and $i=36.2$\degr (drawn line). The observations are binned for clarity, but fitting was done on unbinned data.}
\end{figure}

\section{Analysis and results}
\label{sec:res}
\subsection{Spectral features}
The spectra of \src display a number of absorption lines from the stellar atmosphere of the companion star, but also strong emission lines consistent with H$\alpha$ and H$\beta$ and with He{\sc I} at $\lambda$6678.149 and $\lambda$5875.618.  For each instrument, we combined all the spectra into an average spectrum, after correcting for the Doppler shift of the lines due to the orbital motion (see Section \ref{sec:rvc} for the measurement of the orbital shifts).
The top panel of Figure \ref{fig:spectra} shows the average of the IMACS spectra where the H$\alpha$ and H$\beta$ lines appear as prominent emission features. He{\sc I} in emission at 5875.618 \ang is next to a blend of the Na{\sc I} doublet lines in absorption ($\lambda$5895.92, $\lambda$5889.95) mainly produced by the interstellar medium. The Mg{\sc I} triplet at $\lambda$5167, $\lambda$5172 and $\lambda$5183 is visible, although not resolved. The Ca{\sc I} triplet ($\lambda$6102, $\lambda$6122 and $\lambda$6162) and a blend of stellar lines in absorption at $\lambda$6495 are also present.
\newline
The Doppler-corrected average of the FORS spectra, in the central panel of Figure \ref{fig:spectra}, shows the region of H$\alpha$, with better resolution than that of the IMACS spectrum. He{\sc I} in emission at $\lambda$6678.149 and a forest of absorption lines are visible, in particular from Fe{\sc II} at  $\lambda$6432.65 and $\lambda$6457 and Fe{\sc I} at $\lambda$6430.85. Finally, the bottom panel of Figure \ref{fig:spectra} shows a section of the Doppler-corrected average of the VIMOS spectra, displaying the main features in the red part of the spectrum of \src: the Ca{\sc II} triplet in absorption, at $\lambda$8498, $\lambda$8542 and $\lambda$8662, and a number of emission lines from the Paschen series of hydrogen. A telluric molecular band of water is also present. 
 \newline
In most of the spectra, the H$\alpha$ emission line appears single-peaked but composite, with a main peak overlapping with at least one side wing. Both components are broad, with FWHM of the order of 350-450\,\kms. The equivalent width (EW) of the line is roughly constant around $\sim-13/16$ \ang across the IMACS and FORS spectra, with no dependence on phase. In the MagE spectrum, taken one month later,  it drops significantly to -2.9$\pm0.7$ \ang. The two VIMOS spectra, collected two months earlier than all the other data-sets, suggest a more active state of the source compared to the IMACS and FORS observations. The EW is -133$\pm1$ \ang and -140$\pm1$ \ang in the two spectra. 
The H$\beta$ line has similar EW to H$\alpha$ and an even larger FWHM of $\sim700$\,\kms.   

\subsection{Binary period}
\label{sec:period}
The orbital period of \src was determined using the Mosaic-2 lightcurve, combined with spectroscopic information. The Swope data provide only loose constraints on the system parameters, which we therefore only used as a consistency check of our results.
In order to obtain the best possible accuracy on $P$, we followed three main steps. 
\begin{itemize}
\item An initial Lomb-Scargle periodogram of the Mosaic-2 data revealed an apparent period of 0.12\,d.  Some dispersion was seen when the Mosaic-2 lightcurve was folded on this period indicating that this is the first harmonic. Folding on twice this period yielded an asymmetric double-humped lightcurve (Figure \ref{fig:licus}, {\it top} panel). 
A period of $P$=0.23710(5)\,d was measured by fitting the lightcurve with a sum of two sine waves, with a 2:1 frequency ratio, and amplitudes and relative phases allowed to vary freely.  Error bars were estimated in the usual way from $\chi^2$ totals, after adding an additional error to the Mosaic-2 datapoints to represent unresolved flickering. The Swope data provide a less constrained $P=$0.23(1)\,d, consistent with the Mosaic-2 result.  Unfortunately a joint fit to both data sets was not well constrained due to the difficulty in extrapolating the ephemerides across a year (see below), and possible changes in lightcurve morphology.  
\item The Mosaic-2 $P$ was then used to phase fold the spectroscopic observations and construct an initial radial velocity curve (rvc, the procedure is described in Section \ref{sec:rvc}, for the final orbital period). The rvc looked consistent with orbital motion of the companion star, and was fitted with a sine wave, measuring T$_0$ in between the time of the spectroscopic observations.  We double-checked the measurement of the orbital period leaving it free in the fit, obtaining $P=0.23710(5)$\,d as expected.
\item At last, T$_0$ was compared with the time of the phase 0 in the Mosaic-2 lightcurve, $T_M$. We refined the measurement of $P$ by considering that it must be such that the difference T$_0-$T$_M$ corresponds to an integer number of orbital cycles.  Both the minima in the Mosaic-2 lightcurve potentially correspond to $T_M$, since the $P$ measured above is not accurate enough to unambiguously phase the lightcurve,  collected one year earlier than the spectroscopic observations. Depending on the phasing, we find three possible solutions for $P$. If $T_M$ is at the deepest, primary minimum of the Mosaic-2 lightcurve, then $P=0.237089(3)$\,d, with T$_0-$T$_M=14898\times P$. If $T_M$ corresponds to the secondary minimum, than $P=$0.237169(3)\,d with T$_0-$T$_M=14888\times P$ or $P=0.237009(3)$~d with T$_0-$T$_M=14889\times P$. 
The rvc fitting favors the first solution, as it is the closest to the best-fitting period. The second solution is less likely but, yielding a $\Delta\chi^2$ of 3.5 with respect to the best-fit $\chi^2$, it has a probability of $\sim 20$\% to be correct. The last solution is very unlikely, yielding  $\Delta \chi^2\sim15$. 
\end{itemize}
In conclusion, $P=$\,0.237089(3)\,d is the statistically most likely orbital period for \src, although $P=$\,0.237169(3)\,d can not be ruled out at high confidence. 
\subsection{Radial velocity curve}
\label{sec:rvc}
\begin{figure}
\includegraphics[width=7.5cm, angle=0]{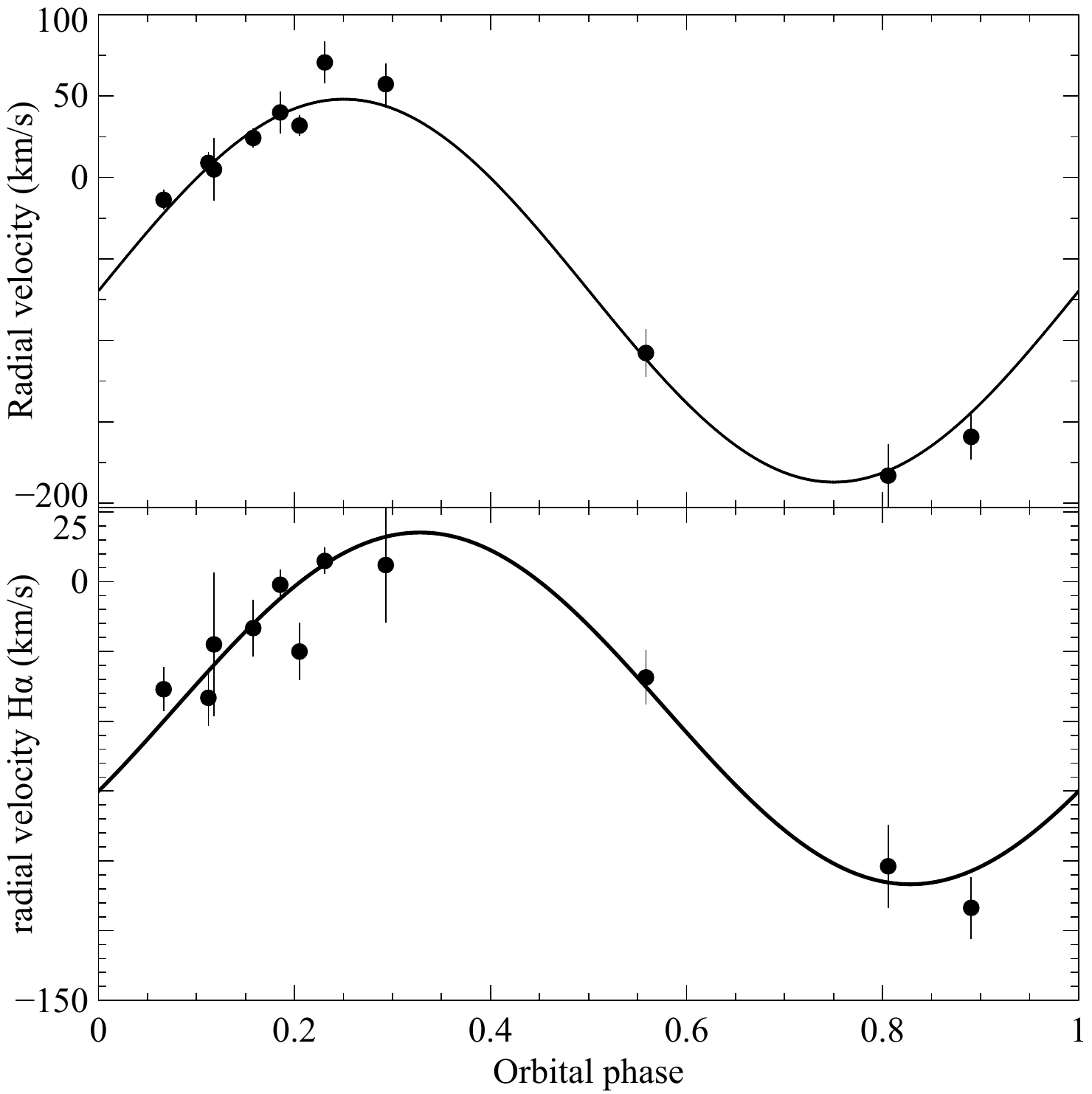} \\
\caption{\label{fig:rvcs} Radial velocity curve measured from the absorption lines ({\it top} panel) and from the peak component of the H$\alpha$ emission line ({\it bottom} panel). The data are fitted with a circular orbit (solid line), increasing the error bars in order to obtain a reduced $\chi^2$ close to 1.}
\end{figure}

The orbital Doppler shifts of the companion star in  \src were measured from the absorption lines in the spectra, by cross-correlating with the K4\,{\sc IV} template star spectrum acquired with FORS. The emission lines were masked. Since we were using observations from different instruments,  the template and each target spectrum were re-binned to the same dispersion and broadened to the same spectral resolution before the cross-correlation. For MagE, the order 10 and the two parts in which the order 9 was split were cross-correlated separately with the FORS template, and the results were averaged in order to reduce the uncertainty. For all the spectra, the uncertainty on the velocity offsets due to the wavelength calibration has been added in quadrature to the velocity errors from the cross-correlation.  
The observations were phase folded with the most likely orbital period from Section \ref{sec:period}, obtaining the rvc shown Figure \ref{fig:rvcs} ({\it top} panel).
 \newline
The rvc was fitted with a circular orbit of the form v($ \phi$)=$\gamma+$K$_{2}\sin(2\pi\phi + \varphi)$, providing a large reduced $\chi^2$ of $\sim13$ (8 degrees of freedom). The uncertainties on the parameters were estimated assuming that the sinusoidal model was correct, and multiplying the errors of the individual velocity shifts by 3 to reach a reduced $\chi^2$ close to 1. T$_0$ was calculated near the middle of the time of the observations and so that $\varphi=0$, according to the convention of phase 0 at inferior conjunction of the companion star. The period was fixed in the fit. The resulting parameter values are: \\
\hspace*{0.5 cm} T$_0=2455743.1892\pm0.0029$~HJD/UTC \\
\hspace*{0.5 cm} K$_2=117\pm6$\,\kms \\
and  $\gamma=-69\pm6$\,\kms, in the rest frame of template star employed for the cross correlation. A systemic radial velocity of 29.8$\pm$0.4\,\kms was found for the latter, by fitting a Gaussian function to the H$\alpha$ absorption line and measuring the offset of the line centroid with respect to its rest-frame wavelength. The systemic velocity for \src is thus \\
\hspace*{0.5 cm} $\gamma=-$39$\pm$6~$\rmn{km\,s^{-1}}$. \\
Figure \ref{fig:rvcs} ({\it bottom} panel) shows the rvc for the H$\alpha$ line, where the orbital velocities were measured through Gaussian fitting to the line peak in each spectrum. The curve was constructed and fitted as we did above. The orbital motion is close to that of the companion star (from the absorption lines), only slightly leading in phase by $\varphi=$ 0.08$\pm$0.03. The fit provides K$_{em}=58\pm6$\,\kms, smaller than the radial velocity semi-amplitude we measured from the absorption lines, suggesting an origin of the H$\alpha$ line towards the inner face of the companion. The \src systemic velocity is $-45\pm4$\,\kms, consistent at the 1$\sigma$ level with the result above. 
Similarly, we measured radial velocity shifts for the H$\alpha$ wing feature. We find variable velocity shifts from $\sim10$ up to $\sim$600\,\kms towards the red or the blue, with no clear dependence on the orbital phase.  

\begin{figure}
\includegraphics[width=8.7cm, angle=0]{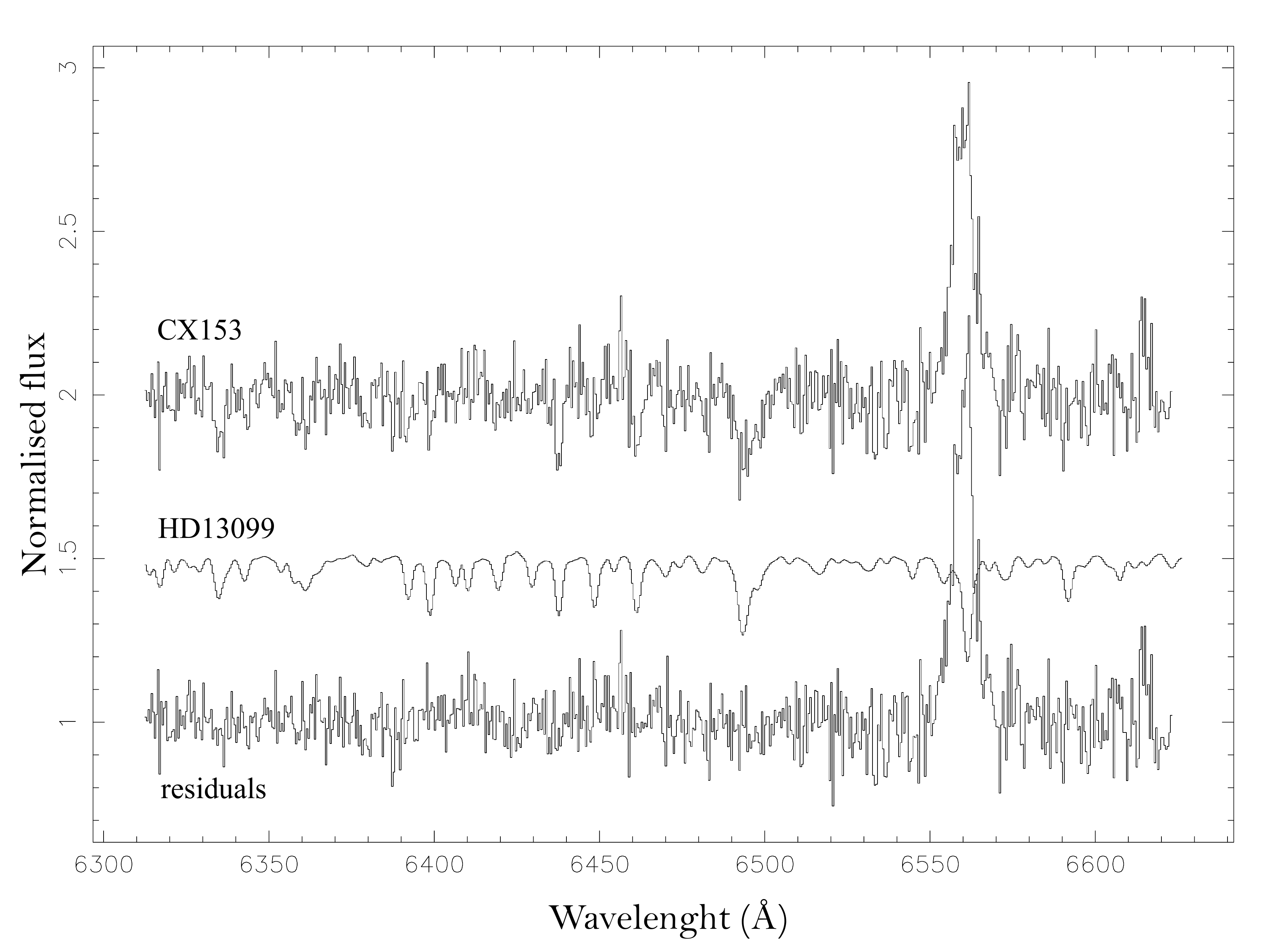} \\
\caption{\label{mage} Optimal subtraction of the spectrum of the stellar template HD13099 from the spectrum of \src, both acquired with MagE, providing the measurement of $v\sin i$. The image shows the region around H$\alpha$ from MagE order 9. The emission lines were masked in the optimal subtraction (see text Section \ref{sec:vsini}). }
\end{figure}

\subsection{Rotational broadening and mass ratio}
\label{sec:vsini}
\noindent The observed FWHM of the absorption lines in the spectra is determined by the intrinsic line width, expected to be dominated by rotational broadening, broadened by the instrumental resolution profile and smeared by the motion of the  companion star during the integration time of one observation. 
In order to measure the intrinsic line width, namely $v\sin i$, we analysed the MagE spectra, which have the best resolution among our data-sets. We considered all the absorption lines in the spectrum (none of which is known to be strongly affected by thermal or pressure broadening), masking the emission lines. The spectral type of the MagE template, K3.5{\sc V}, is likely close to that of the companion star in \src (see Section \ref{sec:sptype}). \citet{Stee07} showed that a small difference in the spectral type of target and template star do not significantly effect the measurement of $v\sin i$ in close binaries. The integration time smearing on the target spectra was taken into account by artificially smearing the template spectrum by $2 \pi T K \cos(2\pi\phi)/P$\,\kms, where $T$ is the duration of one exposure on \src. After that, we broadened the template spectrum with different trial values of  $v\sin i$. For each trial value, the template was subtracted from the target spectrum and a $\chi^2$ test was performed on the residuals. Different scalings of the template are tried to account for possible disc veiling\footnote{{the optimal subtraction is performed using the \sc optsub} command in {\sc molly}. The routine works on normalized spectra, assuming $F_d=F_s-F_t*f$, where $F_s$ is the flux of the target, $F_t$ of the template, $F_d$ is the disc contribution outside the lines and $0<f<1$ is a constant, the optimum factor. The routine finds the $f$ that minimizes the structure in the residuals between $F_d$ and a smooth version of $F_d$ ($<F_d>$), accounting for broad features in the disc spectrum. The disc veiling is $1-f$.}. The broadening that gives the minimum  $\chi^2$ is a measure of the actual $v\sin i$. A value of 0.5 for the limb darkening was assumed, appropriate for a main sequence star as we expect for a $\sim$5 hour orbit. \citet{Stee07} tested the effect of the uncertainty on the limb darkening coefficient on $v\sin i$, showing that it is $\sim1$\,\kms for a difference of 0.25 in the limb darkening. This is negligible with respect to the error on our measurement (see below). Finally, since we subtract template and target spectra acquired with the same instrument, the instrumental resolution profile is not affecting the procedure. 
\newline
The uncertainty on $v\sin i$ was estimated following the Monte Carlo approach used by \citet{Stee07}. We copied the target spectrum 500 times, using a bootstrap technique where the input spectrum is resampled by randomly selecting data points from it. The bootstrapping maintains  the total number of data points in the spectrum. For each bootstrap copy, one value of $v\sin i$ was measured, corresponding to the minimum $\chi^2$ of the residuals as described above. The distribution of $v\sin i$ obtained from the 500 copies is well described by a Gaussian, whose mean and root-mean-square (rms) provides the best-fit $v\sin i$ and its 1$\sigma$ error. Our final measurement of the rotational broadening is the uncertainty-weighted average of the measures from the first and second piece of the MagE order 9, $v\sin i=69\pm7$\,\kms  (Figure \ref{mage}). 
\newline
Given $v\sin i$ and K$_2$ we could calculate the system mass ratio $q=M_1/M_2$ from the relation ${v \sin i \over K_2}\approx0.462q^{1/3}(1+q)^{2/3}$ \citep{Wad88}.  The uncertainty on q was again estimated with a Monte-Carlo simulation, calculating q for 1000 sets of  $v\sin i$ and K$_2$ randomly selected within the 1$\sigma$ uncertainty on the parameters. 
The average of the 1000 simulated measurements provides $q=0.7\pm0.1$, where the uncertainty is the standard deviation of the sample. 


\subsection{Companion star spectral type}
\label{sec:sptype}

In order to constrain the spectral type of the mass donor in \src, we considered a set of 21 high signal-to-noise template spectra of G and K stars of luminosity class {\sc V}, {\sc III} and {\sc IV}, acquired with the William Herschel and the Isaac Newton telescopes between 1992 and 1994. Some of them were previously used for spectral type classification by \citet{Casa96} and \citet{Tor02}. 
We subtracted each template spectrum from the Doppler-corrected average of our FORS observations, performing a $\chi^2$ test on the residuals of the subtraction. The template resulting in the minimum $\chi^2$ provides our best  estimate for the source spectral type. The spectra were shifted to the rest frame of the FORS average spectrum, and degraded to match the FORS sampling and the line broadening in the target spectra. The appropriate broadening (accounting for the difference in resolution between the templates and the FORS spectrum and the line broadening intrinsic to the source) was found in a similar manner as we did for measuring $v\sin i$: different broadened versions of each template were subtracted (through optimal subtraction, accounting for veiling) from the \src spectrum, until the minimum $\chi^2$ was obtained from the fitting of the residuals. As for  $v\sin i$, we assumed a limb darkening of 0.5. 
\newline
The minimum $\chi^2$ is obtained for the K5\,{\sc V} template 61CYGA, with an optimum factor of 0.98$\pm0.01$, indicating a disc veiling of $\sim2$\%. It is reasonable to assume an uncertainty of one spectral type when using this procedure.
\newline
Using the same method decribed in \citet{Rat10} (but adopting standard stellar magnitude in the 2mass J, H and K filters from Mamajek 2011\footnote{http://www.pas.rochester.edu/$\sim$emamajek/ \\ EEM\_dwarf\_UBVIJHK\_colors\_Teff.dat}) we find that the spectral type K5\,{\sc V} is fully consistent with the J, H and K magnitude of \src from VVV, for a reasonable value of the extinction of A$_V=2.0\pm0.6$ (A$_K=0.23\pm0.07$, see Discussion).


\subsection{Ellipsoidal modulations and system inclination}
\label{inc}
The phase folded Mosaic-2 and Swope lightcurves in Figure \ref{fig:licus} show variability between the two data-sets. The Swope lightcurve displays pure ellipsoidal modulations, with equal maxima and the deepest minimum at phase 0.5, while the Mosaic-2 one, collected one year earlier, presents asymmetric maxima. 
\newline
We attempted to determine the system inclination by modeling the lightcurves with the XRbinary program written by E.L. Robinson\footnote{http://pisces.as.utexas.edu/robinson/XRbinary.pdf}. 
The Mosaic-2 Sloan $r^\prime$ and the Swope Gunn $r$ filters were represented by a 5548-6952\,\ang and by a 6181-7212 \,\ang square bandpass,  respectively. In the modeling we assume K$_2$ and $q$ as determined in Sections \ref{sec:rvc} and \ref{sec:vsini}, and that the mass donor is a K5{\sc V} star (see Section \ref{sec:sptype}) with an effective temperature of 4500 K (Mamajek 2011). As we did in Section \ref{sec:period}, the Mosaic-2 data were primely used to constraint the inclination, while the Swope observations was kept for consistency checks only, due to the poorer quality of the data. 
\newline
Depending on the phasing, we fitted the Mosaic-2 lightcurve (Figure \ref{fig:licus}, {\it top} panel) with two models. For the most likely period, the lightcurve has the deepest minimum at phase 0.5 and the highest maximum at phase 0.25. We modeled this with ellipsoidal modulations plus a single (corotating) hot spot on the outer edge of an otherwise non-luminous disc, centred so that it faces earth at phase 0.25 (which is unusual, see Discussion). This provides a reduced $\chi^2$ of $\sim$4 (30 d.o.f.) and an inclination of $32.3\pm1.3\degr$.  
\newline
For the alternative orbital period, the minimum at phase 0.5 is no longer the deepest, which implies heating of the side of the secondary facing the primary. 
We modeled this assuming irradiation from the primary component. On top of ellipsoidal variations, a disc-edge spot, now at phase 0.75, is again included in the model to account for the asymmetric maxima. Since this model depends on the (unknown) disc geometry, two disc versions were tried, with a height to outer radius ratio of 1 to 12 and 1 to 24. 
The resulting system inclination is $36.2\pm1.3\degr$ and $34.5\pm1.3\degr$, respectively.
\newline
The Swope data, modeled with ellipsoidal modulations only, are consistent with all the values of the inclination determined above. Because the Swope observations were taken close to the spectroscopic ones (one night after the last IMACS pointing), and the disc veiling was $\sim$2\% in the spectra, it is safe to assume very little disc contribution to the Swope lightcurve. Its consistency with the inclination estimated from the Mosaic-2 data therefore suggests that the latter are not heavily affected by an accretion disc contribution either (on the effect of a disc contamination on $i$ see \citealt{Can10}).

\subsection{Mass of the stellar components}
\label{masses}
The mass function $f(M_2):  M_1{sin^3i\over (1+q)^2} ={P{K_2}^3\over2\pi G}\approx$ 0.04 of \src is too small to constrain the nature of the primary. However, including our measurements of  $i$ and $q$, we can solve it for all dynamical parameters. For the most probable period (see Section \ref{sec:period}), we obtain M$_1=0.8\pm0.2$\,\msun and M$_2=0.6\pm0.2$\,\msun. The uncertainty was obtained with a Monte-Carlo method, as we did for $q$ in Section \ref{sec:vsini}. 
For the alternative period the inclination is more uncertain and the masses are reduced to M$_1\sim0.6$\,\msun and M$_2\sim0.4$\,\msun.

\subsection{UV counterpart}
We searched for an ultraviolet (UV) counterpart to \src in 900 seconds long observations from the Galaxy Evolution Explorer (GALEX). The closest object to the source is a 2$\sigma$-detection at an angular distance of 2.2\arcsec from our \chan position. Even assuming that the detection was real, the position is not consistent with that of \src (the 1$\sigma$ accuracy of the astrometry of GALEX is 0\farcs4, plus a systematic uncertainty of 0\farcs2 in Dec). The non-detection of a counterpart in GALEX provides an upper limit to the magnitude of  \src of $\sim$22 AB magnitudes in the near UV GALEX filter (at 2500 \ang, limiting magnitude at 3$\sigma$). 


\section{Discussion}
\label{sec:disc}

We have performed phase resolved optical spectroscopy and photometry of the optical counterpart to \src, constraining the system parameters. We found an orbital period of more than 5.6 hours (most likely $P=$\,5.69014(6)\,hours) and an inclination of 32-36\degr (most likely $32.3\pm1.3\degr$). The best-estimated masses of the primary and secondary star are 0.8$\pm0.2$\,\msun and 0.6$\pm$0.2\,\msun, although lower masses are possible depending on the phasing of the Mosaic-2 lightcurve. The latter displays minima consistent with ellipsoidal modulations, but asymmetric maxima (this is often called O'Connel effect, see \citealt{Wils09} for a review). Two values of the period, differing by $\sim$13 seconds, provide an acceptable fit to both the lightcurve and the radial velocity data, causing an ambiguity of half  cycle in the phasing. For the most likely period, we could model the lightcurve with ellipsoidal modulations plus a disc spot accounting for the highest maximum, at phase 0.25. The phasing of the disc spot is unusual though, as a mass transfer stream hitting the accretion disc is expected to produce a hot spot that leads the donor. Alternative effects causing the shape of the lightcurve, consistent with the phasing, are tidal interaction of the companion star with the outer disc (\citealt{Fra02}, although the most commonly observed tidal interaction effect, superhumps, can only happen in systems with $q<0.25$, \citealt{Hyn12}) or starspots reducing the light at one maximum (see, e.g., \citealt{Wils09}). The latter are often invoked to explain to the O'Connel effect, but a realistic model of starspots is hard to produce with lightcurve modeling codes, especially when a single band lightcurve is available. Compared to the disc-spot model, the effect of a starspot model on the system parameters would be to increase the inclination, leading to smaller masses. 
\newline
All our data indicate that the primary is a WD, placing \src in the class of cataclysmic variables (CVs). The mass of the donor is consistent with the spectral type favored by our spectroscopic observations (K5\,{\sc V}, typical mass is $~0.67$\,\msun; \citealt{Dri2000}), or slightly under-massive for the highest inclination scenario. Under-massive companions are often observed in XRBs and CVs as a result of their accretion history (e.g., Her~X--1 and Cyg~X--2, see \citealt{Tau06}, \citealt{Kni06}). 
The de-reddened upper limit to the UV flux that we could obtain from the Galex observations (assuming A$_V$ as derived in section \ref{sec:sptype}) is rather uncertain, due to possible deviation of the extinction law compared to the standard one we used (from \citealt{Car89}). Still, by comparing WD models from \citet{Koe05} with our UV upper limit, we conclude that the non-detection of a UV counterpart is consistent with the presence of a 15000-30000 K WD (in agreement with the WD temperature we find for the irradiation model of the Mosaic-2 lightcurve) while a much hotter object would have been detected.

\subsection{Distance and X--ray luminosity}
By comparing the typical K-band magnitude of a K5{\sc V} star (M$_K=$4.42, Mamajek 2011) with the observed magnitude m$_K=13.76\pm0.03$, and assuming A$_V=2.0\pm0.6$ (A$_K=0.23\pm0.07$) as estimated in Section \ref{sec:sptype}, we obtain the distance  $d=$664$\pm$23\,$\rmn{pc}$ to \src. 
This estimate relies on the assumption that the absolute magnitude of the companion is typical of a normal main sequence star, which is not true for CVs in general \citep{Kni06}. The typical radius of a K5{\sc V} star is 0.72\,\rsun \citep{Dri2000}. Using the mass-to-radius relation found by \citet{Kni06} for secondary stars in long orbital period CVs, we find that our best mass estimate for the donor, 0.6$\pm$0.2\,\msun, yields a radius of $0.62\pm0.07$\,\rsun (this is consistent with the Roche lobe radius of the companion $R_{RL2}=$0.62\,\rsun for M$_1=0.8$ and $q=0.7$). The secondary in \src is then fainter than a typical K5{\sc V} by 0.1 to 0.6 magnitude, allowing a distance of $\sim$500\,$\rmn{pc}$.
Further uncertainty to the distance comes from the assumed spectral type. Allowing the spectral type to be off by one, our observations suggest a likely distance between $\sim400-700$\,$\rmn{pc}$.
\newline
Although seemingly high, the extinction we estimated is consistent with a nearby object. In fact, extinction maps obtained from red clump stars within VVV, with a resolution of 2$^\prime\times$2$^\prime$, indicate E(B-V)=1.67 (A$_V\sim5.1$) in the direction of \src \citep{Gon11}. 
\newline
Converting $A_V$ into an hydrogen column density (following \citealt{Guv09}) yields N$_H=(0.4\pm0.1)\times10^{21}\,\rmn{cm^{-2}}$. Assuming a power-law spectrum with index 1.6,  the 14 counts observed by \chan from \src in the full ACIS range (0.3-8 keV) correspond to an unabsorbed flux of 3.9$\times$10$^{-13}$\,$\rmn{erg\,s^{-1}\,cm^{-2}}$. For $d=$664\,$\rmn{pc}$ the 0.3-8 keV X--ray luminosity of \src is 1.8$\times10^{31}$ \ergsec. Compared with typical CVs luminosities ($\sim10^{31}-10^{34}$\,\ergsec) this is consistent with a low accretion state.

\subsection{The difficult classification of quiescent XRBs}
The low X--ray luminosity, the the lack of clear signatures of ongoing accretion and the little disc veiling in the optical spectra indicate that \src was in a low state during most of our observations. Disc emission is possibly responsible for the wings at the base of the H$\alpha$ line, showing high radial velocities of a few hundred \kms, but the peak of the line traces the donor star (only slightly leading in phase) and can hardly be reconciled with disc motion.
The low state and dominance of the companion over the spectra makes the classification of the system in terms of CV subclasses difficult. An intermediate polar scenario, consistent with the presence of little disc, seems to be ruled out by the lack of typical lines from magnetic CVs in the spectra (such as prominent He{\sc II}, \citealt{Wil89}, \citealt{Schw04}), but even those lines could be absent if the accretion rate is very low. 
The complex H$\alpha$ line resembles nova-like systems such as BB Doradus \citep{Sch12} which do show composite emission features that are not directly from either the secondary star or a remnant disc. Although the dynamics of the line suggests an origin near the tip of the companion star, in fact, the observed FWHM is too broad ($\sim400$\,\kms) to be ascribed to photospheric activity or irradiation of the secondary surface only (although saturation effect might play an important role). \src also shows some similarity with the VY Sculptoris subclass of nova-like CVs (NL), which do show extended low accretion states when there can be hardly any evidence of ongoing accretion (see, e.g. the low state of MV Lyrae in \citealt{Lin05}). A dwarf nova scenario also seems viable, with the source ongoing a small outburst at the time of the Mosaic-2 observations, but with the low accretion rate preventing clear accretion features from showing the classification remains tentative. 
\newline
\src also shows that a little level of variability in a quiescence CV can produce spectral features resembling a quiescent BH or NS. 
Plotting the EW of the Balmer lines detected in the VIMOS spectra against the X--ray to optical flux ratio ($\log(F_X/F_V)\sim-1$, following \citealt{P&R85}) provides outlying values for a CV that are suggestive of a more compact accretor.  The values are consistent with a normal CV if we consider the IMACS spectra instead. 
Moreover, it is worth noticing that reference quantities such as the X--ray luminosity and $F_X/F_V$ are not good indicators of the source type for deeply quiescent systems as they are for actively accreting ones. The typical X--ray luminosity for quiescent NS or BH XRBs ($10^{31}-10^{32}$ \ergsec) is in fact consistent with that of a CV in a low state, and $F_{V}$ is the same for all XRBs when the companion star dominates the optical light.  
In conclusion, in a quiescent system like \src even the nature of the accretor can be mistaken without a complete dynamical study (see also \citealt{Mar94}). This is important to take into account for projects, such as the GBS, that aim to identify new NSs or BHs. 

\section{Conclusion}

We performed a full dynamical study of the GBS source \tmp(\src), finding that the source is a nearby ($400 \lesssim d \lesssim 700\,\rmn{pc}$) long orbital period ($>5$ hours) CV, in a state of low accretion rate. One episode of accretion at higher rate was possibly caught in one of our lightcurves, displaying asymmetric maxima that could be fitted with a disc spot. 
Alternative causes for the asymmetric maxima, such as the presence of starspots on the companion star, are difficult to realistically model. The secondary is likely of spectral type (close to) K5\,{\sc V} and dominates the spectra, which display a disc contribution of $\sim2\%$. The classification of the source among CV subclasses is difficult. A DN (in particular a VY Sculptoris) or NL system in a low state seems likely, although we can not exclude a magnetic WD accreting at a very low rate. The case of \src highlights that CVs in a low accretion state can be hard to distinguish from quiescent XRBs with a NS or BH primary. 

\section*{Acknowledgments} 
\noindent  R.~I.~H and C.~T.~B. acknowledge support from the National Science Foundation under Grant No. AST-0908789.. R.~I.~H. also acknowledges support from NASA/Louisiana Board of Regents grant NNX07AT62A/LEQSF(2007-10) Phase3-02. P.~G.~J acknowledges support from a  VIDI grant from the Netherlands Organisation for Scientific Research. D.~S. acknowledges support from STFC via an Advanced Fellowship and the Warwick Rolling grant.
Tom Marsh is thanked for developing and sharing his packages {\sc Pamela} and {\sc Molly} and E.L. Robinson for his XRbinary code.
We also acknowledge G. Nelemans for useful discussions and J.E.McClintock for his comments.






\bibliographystyle{mn} \bibliography{cx153.bib}

\begin{thebibliography}{34}
\expandafter\ifx\csname natexlab\endcsname\relax\def\natexlab#1{#1}\fi

\bibitem[{{Alard}(1999)}]{Alard:1999a}
{Alard} C., 1999, \aap, 343, 10

\bibitem[{{Alard} \& {Lupton}(1998)}]{Alard:1998a}
{Alard} C., {Lupton} R.~H., 1998, \apj, 503, 325

\bibitem[{{Cantrell} {et~al.}(2010){Cantrell}, {Bailyn}, {Orosz}, {McClintock},
  {Remillard}, {Froning}, {Neilsen}, {Gelino}, \& {Gou}}]{Can10}
{Cantrell} A.~G., {Bailyn} C.~D., {Orosz} J.~A., {McClintock} J.~E.,
  {Remillard} R.~A., {Froning} C.~S., {Neilsen} J., {Gelino} D.~M., {Gou} L.,
  2010, \apj, 710, 1127

\bibitem[{{Cardelli} {et~al.}(1989){Cardelli}, {Clayton}, \& {Mathis}}]{Car89}
{Cardelli} J.~A., {Clayton} G.~C., {Mathis} J.~S., 1989, \apj, 345, 245

\bibitem[{{Casares} {et~al.}(1996){Casares}, {Mouchet}, {Martinez-Pais}, \&
  {Harlaftis}}]{Casa96}
{Casares} J., {Mouchet} M., {Martinez-Pais} I.~G., {Harlaftis} E.~T., 1996,
  \mnras, 282, 182

\bibitem[{{Charles} \& {Coe}(2006)}]{Cha06}
{Charles} P.~A., {Coe} M.~J., 2006, in Cambridge Astrophysics Series, Vol.~39,
  Compact stellar X-ray sources, {Lewin, W., van der Klis, M.}, ed., Cambridege
  Univ. Press, pp. 215--265

\bibitem[{{Deller} {et~al.}(2012){Deller}, {Archibald}, {Brisken},
  {Chatterjee}, {Janssen}, {Kaspi}, {Lorimer}, {Lyne}, {McLaughlin}, {Ransom},
  {Stairs}, \& {Stappers}}]{Del12}
{Deller} A.~T., {Archibald} A.~M., {Brisken} W.~F., {Chatterjee} S., {Janssen}
  G.~H., {Kaspi} V.~M., {Lorimer} D., {Lyne} A.~G., {McLaughlin} M.~A.,
  {Ransom} S., {Stairs} I.~H., {Stappers} B., 2012, ArXiv e-prints

\bibitem[{{Dickey} \& {Lockman}(1990)}]{Dic90}
{Dickey} J.~M., {Lockman} F.~J., 1990, \araa, 28, 215

\bibitem[{{Drilling} \& {Landolt}(2000)}]{Dri2000}
{Drilling} J.~S., {Landolt} A.~U., 2000, in Allen's Astrophysical Quantities,
  {Cox, A.~N.}, ed., New York: AIP Press; Springer, p. 381

\bibitem[{{Frank} {et~al.}(2002){Frank}, {King}, \& {Raine}}]{Fra02}
{Frank} J., {King} A., {Raine} D.~J., 2002, {Accretion Power in Astrophysics:
  Third Edition}. Cambridge, UK: Cambridge University Press

\bibitem[{{Gonzalez} {et~al.}(2011){Gonzalez}, {Rejkuba}, {Zoccali}, {Valenti},
  \& {Minniti}}]{Gon11}
{Gonzalez} O.~A., {Rejkuba} M., {Zoccali} M., {Valenti} E., {Minniti} D., 2011,
  \aap, 534, A3

\bibitem[{{G{\"u}ver} \& {{\"O}zel}(2009)}]{Guv09}
{G{\"u}ver} T., {{\"O}zel} F., 2009, \mnras, 400, 2050

\bibitem[{{Honeycutt} \& {Kafka}(2004)}]{Hon04}
{Honeycutt} R.~K., {Kafka} S., 2004, \aj, 128, 1279

\bibitem[{{Horne}(1986)}]{Hor86}
{Horne} K., 1986, \pasp, 98, 609

\bibitem[{{Hynes}(2012)}]{Hyn12}
{Hynes} R.~I., 2012, to appear the proceedings of the XXI Canary Islands Winter
  School of Astrophysics. Ed. T Shahbaz, CUP, arXiv:1010.5770

\bibitem[{{Jonker} {et~al.}(2011){Jonker}, {Bassa}, {Nelemans}, {Steeghs},
  {Torres}, {Maccarone}, {Hynes}, {Greiss}, {Clem}, {Dieball}, {Mikles},
  {Britt}, {Gossen}, {Collazzi}, {Wijnands}, {In't Zand}, {M{\'e}ndez}, {Rea},
  {Kuulkers}, {Ratti}, {van Haaften}, {Heinke}, {{\"O}zel}, {Groot}, \&
  {Verbunt}}]{Jon11}
{Jonker} P.~G., {Bassa} C.~G., {Nelemans} G., {Steeghs} D., {Torres} M.~A.~P.,
  {Maccarone} T.~J., {Hynes} R.~I., {Greiss} S., {Clem} J., {Dieball} A.,
  {Mikles} V.~J., {Britt} C.~T., {Gossen} L., {Collazzi} A.~C., {Wijnands} R.,
  {In't Zand} J.~J.~M., {M{\'e}ndez} M., {Rea} N., {Kuulkers} E., {Ratti}
  E.~M., {van Haaften} L.~M., {Heinke} C., {{\"O}zel} F., {Groot} P.~J.,
  {Verbunt} F., 2011, \apjs, 194, 18

\bibitem[{{Knigge}(2006)}]{Kni06}
{Knigge} C., 2006, \mnras, 373, 484

\bibitem[{{Koester} {et~al.}(2005){Koester}, {Napiwotzki}, {Voss}, {Homeier},
  \& {Reimers}}]{Koe05}
{Koester} D., {Napiwotzki} R., {Voss} B., {Homeier} D., {Reimers} D., 2005,
  \aap, 439, 317

\bibitem[{{Linnell} {et~al.}(2005){Linnell}, {Szkody}, {G{\"a}nsicke}, {Long},
  {Sion}, {Hoard}, \& {Hubeny}}]{Lin05}
{Linnell} A.~P., {Szkody} P., {G{\"a}nsicke} B., {Long} K.~S., {Sion} E.~M.,
  {Hoard} D.~W., {Hubeny} I., 2005, \apj, 624, 923

\bibitem[{{Marsh} {et~al.}(1994){Marsh}, {Robinson}, \& {Wood}}]{Mar94}
{Marsh} T.~R., {Robinson} E.~L., {Wood} J.~H., 1994, \mnras, 266, 137

\bibitem[{{Osterbrock} {et~al.}(1996){Osterbrock}, {Fulbright}, {Martel},
  {Keane}, {Trager}, \& {Basri}}]{Ost96}
{Osterbrock} D.~E., {Fulbright} J.~P., {Martel} A.~R., {Keane} M.~J., {Trager}
  S.~C., {Basri} G., 1996, \pasp, 108, 277

\bibitem[{{Patterson} \& {Raymond}(1985)}]{P&R85}
{Patterson} J., {Raymond} J.~C., 1985, \apj, 292, 535

\bibitem[{{Ratti} {et~al.}(2010){Ratti}, {Bassa}, {Torres}, {Kuiper},
  {Miller-Jones}, \& {Jonker}}]{Rat10}
{Ratti} E.~M., {Bassa} C.~G., {Torres} M.~A.~P., {Kuiper} L., {Miller-Jones}
  J.~C.~A., {Jonker} P.~G., 2010, \mnras, 408, 1866

\bibitem[{{Schmidtobreick} {et~al.}(2012){Schmidtobreick},
  {Rodr{\'{\i}}guez-Gil}, {Long}, {G{\"a}nsicke}, {Tappert}, \&
  {Torres}}]{Sch12}
{Schmidtobreick} L., {Rodr{\'{\i}}guez-Gil} P., {Long} K.~S., {G{\"a}nsicke}
  B.~T., {Tappert} C., {Torres} M.~A.~P., 2012, \mnras, 422, 731

\bibitem[{{Schwarz} {et~al.}(2004){Schwarz}, {Schwope}, {Staude}, {Urrutia},
  {Rau}, \& {Hasinger}}]{Schw04}
{Schwarz} R., {Schwope} A.~D., {Staude} A., {Urrutia} T., {Rau} A., {Hasinger}
  G., 2004, in Astronomical Society of the Pacific Conference Series, Vol. 315,
  IAU Colloq. 190: Magnetic Cataclysmic Variables, {Vrielmann} S., {Cropper}
  M., eds., p. 230

\bibitem[{{Shaw}(2009)}]{Shaw:2009a}
{Shaw} R.~A., 2009, NOAO Data Handbook (Version 1.1). Tucson: National Optical
  Astronomy Observatory

\bibitem[{{Steeghs} \& {Jonker}(2007)}]{Stee07}
{Steeghs} D., {Jonker} P.~G., 2007, \apjl, 669, L85

\bibitem[{{Stetson}(1987)}]{Stetson:1987a}
{Stetson} P.~B., 1987, \pasp, 99, 191

\bibitem[{{Tauris} \& {van den Heuvel}(2006)}]{Tau06}
{Tauris} T.~M., {van den Heuvel} E.~P.~J., 2006, {Formation and evolution of
  compact stellar X-ray sources}, Cambridge University Press, pp. 623--665

\bibitem[{{Tokunaga}(2000)}]{Tok00}
{Tokunaga} A.~T., 2000, in Allen's Astrophysical Quantities, {Cox, A.~N.}, ed.,
  Springer-Verlag (New York), p. 143

\bibitem[{{Torres} {et~al.}(2002){Torres}, {Casares}, {Mart{\'{\i}}nez-Pais},
  \& {Charles}}]{Tor02}
{Torres} M.~A.~P., {Casares} J., {Mart{\'{\i}}nez-Pais} I.~G., {Charles} P.~A.,
  2002, \mnras, 334, 233

\bibitem[{{Wade} \& {Horne}(1988)}]{Wad88}
{Wade} R.~A., {Horne} K., 1988, \apj, 324, 411

\bibitem[{{Williams}(1989)}]{Wil89}
{Williams} R.~E., 1989, \aj, 97, 1752

\bibitem[{{Wilsey} \& {Beaky}(2009)}]{Wils09}
{Wilsey} N.~J., {Beaky} M.~M., 2009, Society for Astronomical Sciences Annual
  Symposium, 28, 107

\end{thebibliography}

\end{document}